\documentclass[a4paper,twoside]{article}
\newcommand{\affil}[1]{$^{\rm #1}$}
\usepackage[authoryear]{natbib}
\bibpunct{(}{)}{;}{a}{}{,}
\usepackage{graphicx}
\def\reference{\parskip 0pt\par\noindent\hangindent 0.5 truecm}%

\date{} 
\title{\large\bf\flushleft The Chemical Composition of Alpha Centauri A: Strong
Lines and the ABO Theory of Collisional Line Broadening}
\author{\parbox{\textwidth}{\flushleft
\vspace{-0.5cm}
{\it M. T. Doyle\affil{A,C}, B. J. O'Mara (deceased\affil{1}), J. E. Ross\affil{A},and M. S. Bessell\affil{B}}\\
\vspace{0.4cm}
{\small \affil{A}\,Physical Department, University of Queensland,
St. Lucia, Qld 4072, Australia}\\
{\small \affil{B}\,RSAA, ANU, Mount Stromlo, Cotter Road, Weston Creek,
ACT 2611, Australia}\\
{\small \affil{C}\,Email: mtdoyle@physics.uq.edu.au}}}
\begin{document}
\twocolumn[
\begin{changemargin}{.8cm}{.5cm}
\begin{minipage}{.9\textwidth}
\vspace{-1cm}
\maketitle
%
%
\small{\bf Abstract}\
The mean abundances of Mg, Si, Ca, Ti, Cr and Fe based on both strong
and weak lines of Alpha Centauri A are determined by matching the
observed line profiles with those synthesized from stellar atmospheric
models and comparing these results with a similar analysis for the
Sun. There is good agreement between the abundances from strong and
weak lines. 

Strong lines should generally be an excellent indicator of abundance
and far easier to measure than the weak lines normally used. Until
the development of the Anstee, Barklem and O'Mara theory for collisional
line broadening, the uncertainty in the value of the damping constant
prevented strong lines being used for abundance determinations other
than in close differential analyses.

We found that Alpha Centauri A has a mean overabundance of 0.12$\pm$0.06
dex compared to solar mean abundances. This result agrees remarkably
well with previous studies that did not use strong lines or the Anstee,
Barklem and O'Mara theory for collisional line broadening. Our result
support the conclusion that reliable abundances can be derived from
strong lines provided this new theory for line broadening is used
to calculate the van der Waal's damping.


\medskip
\medskip
\end{minipage}
\end{changemargin}
]
\small

\section{Introduction}

The Alpha Centauri system has a special fascination for astrophysicists
because it is the closest stellar system and its principle component,
Alpha Centauri A ($\alpha$ Cen A), has a spectral type very similar
to the Sun, G2V. $\alpha$ Cen A is also one of the brightest stars
in the sky enabling spectra with a high spectral resolution and high
signal to noise to be obtained. Most past analyses have concluded
that the metal abundance of $\alpha$ Cen A is greater than that of
the Sun. Analyses, such as Furenlid \& Meylan (1990) covering 26 elements
from 500 lines and Neuforge-Verheecke \& Magain (1997) investigating
17 elements, concluded that the average metal overabundance compared
to the Sun is 0.12$\pm$(0.02-0.04) dex and 0.24 dex respectively.
Another study, Chmielewski et al. (1992), concluded that $\alpha$
Cen A can be classed as a super metal rich star. 

The importance of the present study is the use of the Anstee, Barklem,
and O'Mara%
\footnote{Jim O'Mara was the primary supervisor for this project. Sadly he died
suddenly in April 2002 during a trip to Italy and France before this
work was finished.%
} theory (ABO theory, Barklem et al. (1998a) ) to determine
van der Waal's damping (VDW) for collisional line broadening for $\alpha$
Cen A. The ABO theory provides precise theoretical damping constants
(as demonstrated in recent results for the analysis of strong lines
in the solar spectrum by Allende Prieto et al.(2001)), which enables
the use of strong lines for which reliable laboratory f-values exist.
Strong line wings are also relatively insensitive to the effects of
turbulence in the atmosphere. Thus strong line wings, together with
the ABO theory, may be used as reliable abundance indicators for elements
where such strong lines exist.

The choice of lines for this project follows Allende Prieto et al.(2001)
investigation for the Sun and includes weak neutral, weak ionised,
and strong lines for six elements, Mg, Ca, Si, Ti, Cr, and Fe. 

We employ high quality coude echelle CCD spectra observed using the
74 inch telescope at Mt. Stromlo Observatory. The mass, distance,
luminosity, and colours of $\alpha$ Cen A are used to determine its
effective temperature, $T_{eff}$, and surface gravity, log($g_{s}$)
necessary to customise existing atmospheric models to match $\alpha$
Cen A. The model is based on Kurucz solar models, (Kurucz 1979). The
customised model, along with relevant atomic data, are used to synthesize
line profiles for $\alpha$ Cen A. Abundances are determined by matching
the observed profiles, following determination of turbulence parameters.
Steps are taken to verify the $\alpha$ Cen A model used. A solar
model based on the Holweger-M\"uller model (Holweger \& M\"uller
1974) is used to synthesize line profiles which are matched to the
observational line profiles from the Jungfraujoch Atlas (Delbouille
\& Roland 1995) to determine the solar mean abundance. The solar
and $\alpha$ Cen A mean abundances are compared to find the mean
under- or overabundance for $\alpha$ Cen A.

\section{ABO Collisional Line Broadening Theory }

Collisional or VDW broadening is broadening resulting from the collision
of atoms in the photosphere. It is especially important in cool stars
such as our Sun and $\alpha$ Cen A that have predominately neutral
hydrogen photospheres. This collisional broadening produces a Lorentzian
line profile. The damping constant for collisional broadening is $\Gamma_{coll}$
and is included when synthesizing the line profiles. 

As discussed in Barklem et al. (1998a), up until the 1970's, formulations
of VDW broadening developed by Lindholm, Foley, and \"Unsold were
used and it was widely held that a better theory was needed. These
theories deal with VDW interactions between perturbing hydrogen atoms
and the absorbing atom. Although the term van der Waal's broadening
is still used today, it is really a misnomer as the actual line broadening
theory is much more complex. 

K.A. Brueckner (Brueckner 1971) introduced a perturbation theory formulation
that involved long range interaction where the electron exchange could
be neglected but not the overlap in the atomic charge distribution,
a point taken up in O'Mara (1976). O'Mara's work deals with collision
broadening theory and draws on work from many sources to develop the
beginning of what has come to be known as the ABO theory of line broadening.
This theory has been further developed by Anstee, Barkelm and O'Mara
(Anstee \& O'Mara 1991, 1995; Barklem \& O'Mara 1997; Barklem et al.
1998b). Further development of the ABO theory and its relevances to
solar and late-type star abundances is on-going with a code available
on the world wide web (Barklem et al. 1998a) to calculate VDW.

\section{Observation and Data Reduction}

Our spectra are taken on three separate observational runs in 1996
June/July and 2001 May on the 74 inch telescope at Mt Stromlo Observatory.
The 120 inch focal length coude camera is used with a 31.6 groove/mm
echelle, cross dispersed with a 150 lines/mm grating. Several different
wavelength settings of the echelle grating and cross-disperser gratings
are used to obtain almost complete wavelength coverage from 4000-8000
$\AA$. 

The signal to noise ratio (S/N) varied with order across the CCD because
of vignetting resulting from the cross-disperser not being near a
pupil. The CCD has a gain of 2 electrons per ADU and most of the exposures
are aimed at about 60000 electrons maximum. As the data are 60000-180000
electrons per resolution element, the nominal S/N ratio is 200-400.
The actual resolution was 125000 estimated by measuring the width
(FWHM) of the line at wavelength 8252.379 $\AA$ from the thorium arc spectrum:\begin{equation}
R=\frac{\lambda}{\Delta\lambda}\label{Eq-Resolution}\end{equation}

which corresponds to a velocity of 2.392 km s$^{-1}$. 

The reduction process includes cleaning the raw spectra of cosmic
rays and flat fielding. The spectral orders are extracted and any
scattered light between the orders is removed. Extracted spectra from
a nearby ``smooth spectrum'' star $\beta$ Centauri, are divided through
the spectra of the target star, $\alpha$ Cen A to eliminate telluric
(Earth's atmosphere) lines. An Th arc spectrum is used for wavelength
calibration purposes. The spectra are smoothed to reduce noise and
the continuum level is flattened. The wavelengths are also corrected
for the radial velocity of $\alpha$ Cen A.

\section{The Atmospheric Models and Line Profiles}

Line profiles are synthesised for each line of interest and used to
determine the abundances by fitting to the observed line profiles.
Parameters such as macro and microturbulence, VDW damping, energy
levels of the transitions, log gf values, and starting abundances
are input to enable this direct matching process. 

Two models and subsequent line profiles are used for $\alpha$ Cen
A. Model AK is an interpolated Kurucz model (Kurucz 1979) grid; model
AH is a scaled Holweger \& M\"uller solar model (Holweger \& M\"uller
1974). Both models use the calculated values for $\alpha$ Cen A's
effective temperature and surface gravity. These two models are compared
to verify the validity of using the interpolated Kurucz model in this
project.

The third model SH is the Holweger \& M\"uller model (Holweger \&
M\"uller 1974). This is used to synthesize line profiles to fit observed
solar data obtained from the Jungfraujoch Atlas.

The initial models based on published solar abundances, are used to
compute line profiles to be compared with observations. If the line
does not fit, the abundance of the element is adjusted and new number
densities, opacities, and pressures are computed for a second iteration.
This process is iterated to convergence yielding an abundance for
each line. The temperature structure, $T_{log\,\tau}$, of the initial
atmosphere is not adjusted.

\subsection{Input Parameters and Calculations for the $\alpha$ Cen A Model Atmosphere
and Synthesised Line Profiles}

\subsubsection{Effective Temperature and Surface Gravity }

The parameters, effective temperature, $T_{eff}$, and surface gravity,
log($g_{s}$), used to calculate the atmospheric model are very important. 

In Table \ref{Table:ParametersUsed} we list the measurements and
derived data for $\alpha$ Cen A and the Sun that are used to determine
the surface gravities and effective temperatures and are computed
using the following equations:

\begin{equation}
\log\, g_{s}=\log\,\left(\frac{Gm}{r^{2}}\right)\label{Eq-LogG}\end{equation}

\begin{equation}
T_{eff}=\left(\frac{L}{4\pi r^{2}\sigma}\right)^{\frac{1}{4}}\label{Eq:EffTemp1}\end{equation}

The derived values are in close agreement with those researched by
other authors and are listed in Table \ref{Table:Comparing-parameters}.
For $\alpha$ Cen A the resulting values are $T_{eff}$=5784$\pm$5K
and log($g_{s}$)=4.28$\pm$0.01.

%
\begin{table*}

\caption{{\footnotesize \label{Table:ParametersUsed}Researched and Calculated{*}
values used to customise the model atmosphere and produce synthesised
line profiles for $\alpha$ Cen A's and the Sun}}

\begin{center}{\footnotesize }\begin{tabular}{p{2cm}p{2.8cm}p{1.8cm}p{2.8cm}p{1.7cm}}
\hline 
{\footnotesize Parameters}&
{\footnotesize $\alpha$ Cen A}&
{\footnotesize Source}&
{\footnotesize Sun}&
{\footnotesize Source}\tabularnewline
\hline
{\footnotesize Angular Diameter (arc sec)}&
{\footnotesize $\left(86.2\pm2.3\right)\times10^{-4}$$^{(\textrm{1})}$ }&
{\footnotesize Absolute IR photometry}&
&
\tabularnewline
{\footnotesize Parallax (arc sec)}&
{\footnotesize 0.74212$\pm$0.0014$^{(\textrm{2})}$}&
{\footnotesize Hipparcos Catalogue}&
&
\tabularnewline
{\footnotesize Apparent Magnitude V}&
{\footnotesize -0.01$\pm$0.006$^{(\textrm{2})}$}&
{\footnotesize Hipparcos Catalogue}&
{\footnotesize -26.74$\pm$0.06$^{(3)}$ }&
{\footnotesize See Reference}\tabularnewline
{\footnotesize M$_{v}$}&
{\footnotesize 4.35$\pm$0.006{*}}&
{\footnotesize Section \ref{SubSec:InputData4Calc}}&
{\footnotesize 4.83$\pm$0.002{*}}&
{\footnotesize Section \ref{SubSec:InputData4Calc}}\tabularnewline
{\footnotesize Distance Modulus}&
{\footnotesize -4.36$\pm0.008${*}}&
{\footnotesize Section \ref{SubSec:InputData4Calc}}&
{\footnotesize -31.57$\pm$0.06{*}}&
{\footnotesize Section \ref{SubSec:InputData4Calc}}\tabularnewline
{\footnotesize Bolometric Correction}&
{\footnotesize -0.07$\pm$0.01$^{(\textrm{4})}$}&
{\footnotesize Observational \& synthetic V band spectra}&
{\footnotesize -0.07$\pm$0.01$^{(\textrm{4})}$}&
{\footnotesize Observational \& synthetic V band spectra}\tabularnewline
{\footnotesize Bolometric Magnitude}&
{\footnotesize 4.2755$\pm$0.01{*}}&
{\footnotesize Calculated}&
{\footnotesize 4.7620$\pm$0.01{*}}&
{\footnotesize Section \ref{SubSec:InputData4Calc}}\tabularnewline
{\footnotesize Mass (kg)}&
{\footnotesize (2.1591$\pm$0.019)$\times10^{30}${*}}&
{\footnotesize Section \ref{SubSec:InputData4Calc}}&
{\footnotesize 1.99$\times10^{30}$$^{(\textrm{3})}$$^{\diamond}$}&
{\footnotesize See Reference}\tabularnewline
{\footnotesize Mass$_{\odot}$}&
{\footnotesize 1.085$\pm$0.01$^{(5)}$}&
{\footnotesize From Models}&
{\footnotesize 1}&
\tabularnewline
{\footnotesize Radius (m)}&
 {\footnotesize (8.6879$\pm0.017$)$\times10^{8}${*}}&
{\footnotesize Section \ref{SubSec:InputData4Calc}}&
{\footnotesize (6.96$\pm$0.00026)$\times10^{8}$$^{(\textrm{3})}$}&
{\footnotesize See Reference}\tabularnewline
{\footnotesize Radius$_{\odot}$}&
 {\footnotesize 1.2483$\pm$0.0024{*}}&
{\footnotesize Section \ref{SubSec:InputData4Calc}}&
{\footnotesize 1}&
\tabularnewline
{\footnotesize Distance (m)}&
{\footnotesize 4.1578$\pm$0.0078$\times10^{16}${*}}&
{\footnotesize Section \ref{SubSec:InputData4Calc}}&
{\footnotesize 1.496$\times10^{11}$ $^{(\textrm{3})}$$^{\diamond}$}&
{\footnotesize See Reference}\tabularnewline
{\footnotesize Distance (Parsecs)}&
 {\footnotesize 1.3456$\pm$0.0025{*}}&
{\footnotesize Section \ref{SubSec:InputData4Calc}}&
{\footnotesize 4.848$\times10^{-6}${*} $^{\diamond}$}&
{\footnotesize Section \ref{SubSec:InputData4Calc}}\tabularnewline
{\footnotesize Luminosity (W)}&
{\footnotesize 6.0205$\pm$1$\times10^{26}${*}}&
{\footnotesize Section \ref{SubSec:InputData4Calc}}&
{\footnotesize 3.846$\times10^{26}$ $^{(\textrm{3})}$$^{\diamond}$}&
{\footnotesize See Reference}\tabularnewline
{\footnotesize $\frac{L}{L_{\odot}}$}&
{\footnotesize 1.57$\pm$0.3{*}}&
{\footnotesize Section \ref{SubSec:InputData4Calc}}&
{\footnotesize 1}&
\tabularnewline
{\footnotesize Surface Gravity (ms$^{-2}$)}&
{\footnotesize (1.901$\pm$0.019)$\times10^{4}${*}}&
{\footnotesize Section \ref{SubSec:InputData4Calc}}&
{\footnotesize (2.74$\pm$0.0002)$\times10^{4}${*}}&
{\footnotesize Section \ref{SubSec:InputData4Calc}}\tabularnewline
{\footnotesize Log g$_{s}$}&
 {\footnotesize 4.28$\pm$0.01{*}}&
{\footnotesize Section \ref{SubSec:InputData4Calc}}&
{\footnotesize 4.44$\pm$0.00003 {*}}&
{\footnotesize Section \ref{SubSec:InputData4Calc}}\tabularnewline
{\footnotesize Effective Temperature (K)}&
{\footnotesize 5784.3$\pm$5.5{*}}&
{\footnotesize Section \ref{SubSec:InputData4Calc}}&
{\footnotesize 5778$\pm$1$^{(\textrm{3})}$}&
{\footnotesize See Reference}\tabularnewline
\hline
\end{tabular}\end{center}{\footnotesize \par}

{\footnotesize $^{(\textrm{1})}$Blackwell \& Shallis 1977; $^{(\textrm{2})}$Perryman
1997; $^{(\textrm{3})}$Ahrens 1995;
$^{(4)}$Bessell et al. 1998; $^{(\textrm{5})}$Demarque et al. 1986,
{*} calculated, $^{\diamond}$ error insignificant}
\end{table*}

\begin{table*}

\caption{{\footnotesize \label{Table:Comparing-parameters}Comparing parameters
for $\alpha$ Cen A from reviewed and this paper}}

\begin{flushleft}{\footnotesize }\begin{tabular}{ccccccccc}
\hline 
{\footnotesize Ref.}&
{\footnotesize T$_{eff}$(K)}&
{\footnotesize log(g$_{\textrm{s}}$)}&
{\footnotesize $\xi$(km s$^{-1}$)}&
{\footnotesize M$_{v}$}&
\textbf{\footnotesize $\frac{L}{L_{\odot}}$}&
\textbf{\footnotesize $\frac{r}{r_{\odot}}$}&
\textbf{\footnotesize $\frac{m}{m_{\odot}}$}&
{\footnotesize $\Delta\log(A)$(dex)}\tabularnewline
\hline
{\footnotesize $^{\textrm{1}}$}&
{\footnotesize 5784.3$\pm$5.5}&
{\footnotesize 4.28$\pm$0.01}&
{\footnotesize 1}&
{\footnotesize 4.35$\pm$0.006}&
{\footnotesize 1.57$\pm$0.3}&
{\footnotesize 1.25$\pm$0.0024}&
{\footnotesize 1.085$\pm$0.01}&
{\footnotesize 0.12$\pm$0.06}\tabularnewline
{\footnotesize $^{2}$}&
{\footnotesize 5770$\pm$20 }&
&
&
&
&
{\footnotesize 1.23}&
&
\tabularnewline
{\footnotesize $^{3}$}&
{\footnotesize 5710$\pm$25}&
{\footnotesize 4.0$\pm$0.2}&
{\footnotesize 1.0$\pm$0.2}&
{\footnotesize 4.38}&
&
{\footnotesize 1.26}&
{\footnotesize 1.085}&
{\footnotesize 0.12$\pm0.06$}\tabularnewline
{\footnotesize $^{4}$}&
{\footnotesize 5765$\pm$50}&
&
&
&
{\footnotesize 1.53}&
&
{\footnotesize 1.085}&
{\footnotesize 0.25$\pm$0.02}\tabularnewline
{\footnotesize $^{5}$}&
{\footnotesize 5800$\pm$20}&
{\footnotesize 4.31$\pm0.02$}&
{\footnotesize 1}&
{\footnotesize 4.374$\pm$0.01}&
&
&
{\footnotesize 1.085}&
\tabularnewline
{\footnotesize $^{6}$}&
{\footnotesize 5830$\pm30$}&
{\footnotesize 4.30$\pm$0.03}&
{\footnotesize 1.09$\pm0.11$}&
&
{\footnotesize 1.53}&
&
{\footnotesize 1.085$\pm0.01$}&
\tabularnewline
\hline
\end{tabular}\end{flushleft}{\footnotesize \par}

{\footnotesize $^{(\textrm{1})}$This paper; $^{(\textrm{2})}$Soderblom
1986; $^{(\textrm{3})}$Furenlid \& Meylan 1990; $^{(4)}$Noels et
al. 1991; $^{(\textrm{5})}$Chmielewski et al. 1992; $^{(\textrm{6})}$Neuforge-Verheecke
\& Magain 1997.}
\end{table*}

\subsubsection{Input Data for Line Profile Synthesis \label{SubSec:InputData4Calc}}

The input data required for line synthesis are the energy levels,
log(gf), and VDW parameters taken from Allende Prieto et al. (2001)
(Tables \ref{Tab-FinishedAbund1} \& \ref{Tab-FinishedAbund2}), and
the starting solar abundances from Grevesse \& Sauval (1998) (Table
\ref{Tab-Mean A & Del A}).

The parallax ($\pi$) and angular diameter ($\theta$) are used to
calculate $\alpha$ Cen A's radius:

\begin{flushleft}\begin{equation}
d=\frac{(d_{\odot})}{\pi}\label{Eq:diameter}\end{equation}
\end{flushleft}

\begin{equation}
r=\frac{[d(\theta/3600)(\pi/180)]}{2}\label{Eq:radius}\end{equation}

The apparent magnitude is used to calculate the effective temperature
with the following steps:\begin{equation}
M_{v}=V-5\log\frac{d}{10}\label{Eq-AbMg}\end{equation}

\begin{equation}
M_{bol}=M_{v}+BC\label{Eq-BolMag}\end{equation}

\begin{equation}
L=L_{\odot}10^{\left(\frac{M_{Bol\odot}-M_{Bol}}{2.5}\right)}\label{Eq-LumACA}\end{equation}

\begin{equation}
T_{eff}=\left(\frac{L}{4\pi r^{2}\sigma}\right)^{\frac{1}{4}}\label{Eq-Teff}\end{equation}

\subsection{Non-thermal Broadening Parameters}

The non-thermal motions or turbulence that occur on a large scale
compared with optical depth, is macroturbulence, and on the small
scale is microturbulence.

\paragraph*{Macroturbulence }

Macroturbulence will broaden the line profile but does not affect
the line strength (equivalent width). The effects of stellar rotation
(small) and instrumental profile are included as enhancements to the
macroturbulence.

Several single lines and one blend of lines are chosen to determine
the effective macroturbulence. Various values are tested on each single
line to see if a common value can be used to match the general shape
of the observed line. When the value for the effective macroturbulence
is determined, that value is used on the blended lines with no affect
on the equivalent width of the synthesised line profile. 

The $\alpha$ Cen A value for the effective macroturbulence is determined
to be 3.3 km s$^{-1}$. This includes 2.4 km s$^{-1}$ due to instrumental
effects and an atmospheric macroturbulence of 2.3 km s$^{-1}$. This
indicates that most of the observed line broadening effects near the
line core are from the large scale motions in the photosphere. Natural,
Stark, and rotational broadening are insignificant for the lines chosen
for this study. 

The effective macroturbulence to match the solar model AH's line profiles
to the observed solar data is 1.6 km s$^{-1}$.

\paragraph*{Microturbulence }

Values for the microturbulence are determined by using weak neutral,
weak ionised, and strong Fe lines. The values fitted are 1.08$\pm$0.2
km s$^{-1}$ for $\alpha$ Cen A with the AK model and 0.85 km s$^{-1}$
for the Sun using the SH model.

\subsection{Validity of the Model}

To judge whether the interpolated model AK is valid, comparison with
an empirical model based on observations is useful. 

The equivalent widths and profiles from model AK are compared to those
produced with model AH a scaled solar empirical Holweger \& M\"uller
model (Holweger \& M\"uller 1974). 

Eight lines are chosen and a good match of line profiles and equivalent
widths are obtained.

\begin{flushleft}%
\begin{figure}[h]
\begin{flushleft}\includegraphics[%
  width=8cm,
  height=7cm]{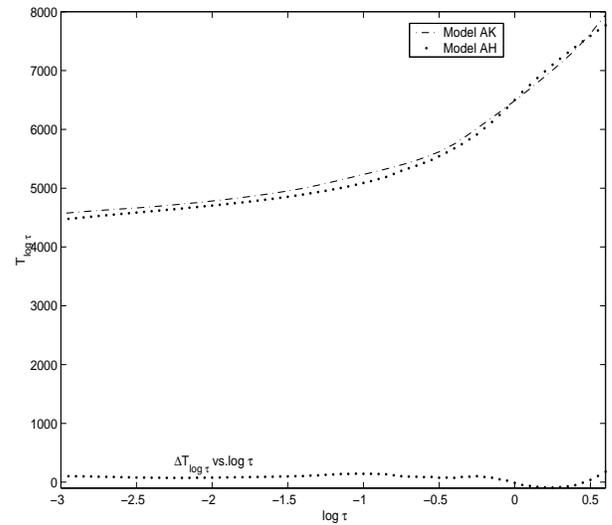}\end{flushleft}

\caption{{\footnotesize Temperature vs. Optical Depth. Comparison of the AK
and AH Models to test for for validity.\label{Fig-ModelComparison} }}
\end{figure}
\end{flushleft}

\begin{flushleft}%
\begin{figure}[h]
\begin{flushleft}\includegraphics[%
  width=10cm,
  height=7cm]{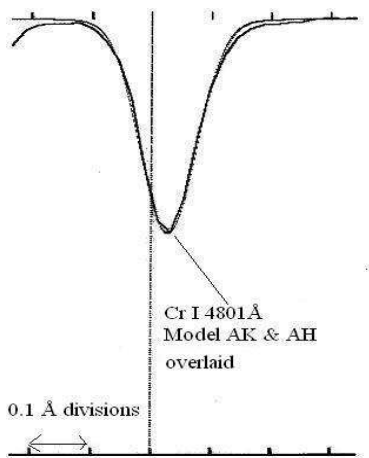}\end{flushleft}

\caption{{\footnotesize \label{Fig:ModelAK&AHOverlay}Model AK \& AH line
profiles overlaid for Cr}{\scriptsize I} {\footnotesize 4801 $\AA$}}
\end{figure}
\end{flushleft}

Another method for comparing the validity of model AK is to plot $T_{log\,\tau}$
vs. log$\tau_{o}$ for both models. As can be seen from Figure \ref{Fig-ModelComparison}
there is close agreement for log$\tau_{o}$ -3-+0.6. This is the range
of optical depths that are covered in this research. A direct comparison
can be seen in Figure \ref{Fig:ModelAK&AHOverlay} where a strong
line, Cr{\small I} at 4801$\AA$, for both models have been
overlaid and match exactly. 

\section{Analysis}

\subsection{Determining the Mean Abundance}

Fitting the shape of the synthesised line profiles from model AK against
that of the observations is straightforward for most of the lines.

For all lines, care is taken to ensure that the area of the synthesised
and observed line profiles are equivalent, allowing for differences
in the exact matching of the line profile's core and wings. For two
very weak lines and all the strong lines, extra care is needed.

In the case of the strong lines, the core and wings cannot be matched
simultaneously. In this case, the wings are matched, as per the ABO
theory, and the core of the model's line profile is extended as low
as possible. Care is taken that this technique is followed in both
model's line profiles for those lines affected.

\begin{figure*}
\begin{flushleft}\includegraphics[%
  width=7cm,
  height=7cm]{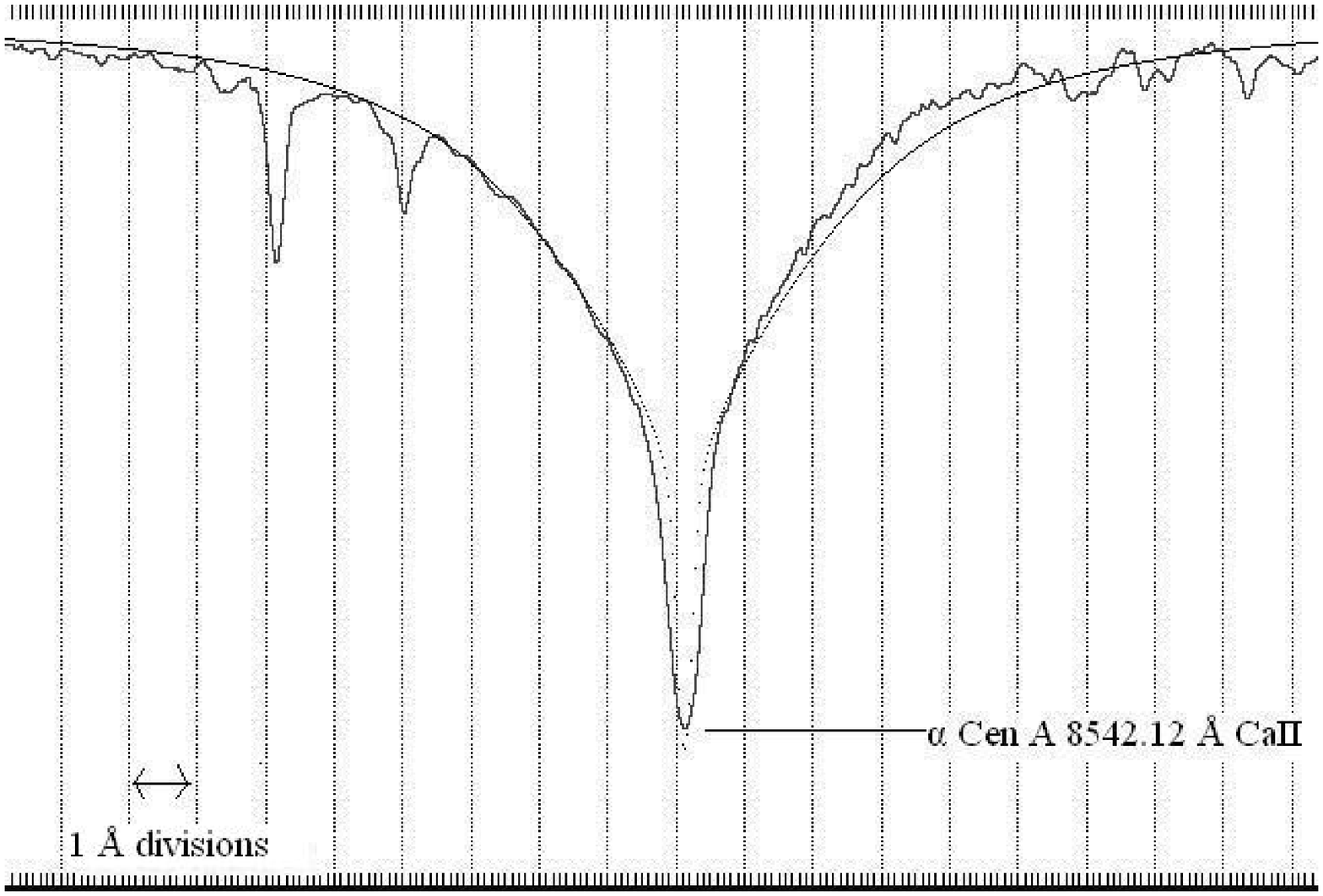}
 \ \ \ \includegraphics[%
  width=7cm,
  height=7cm]{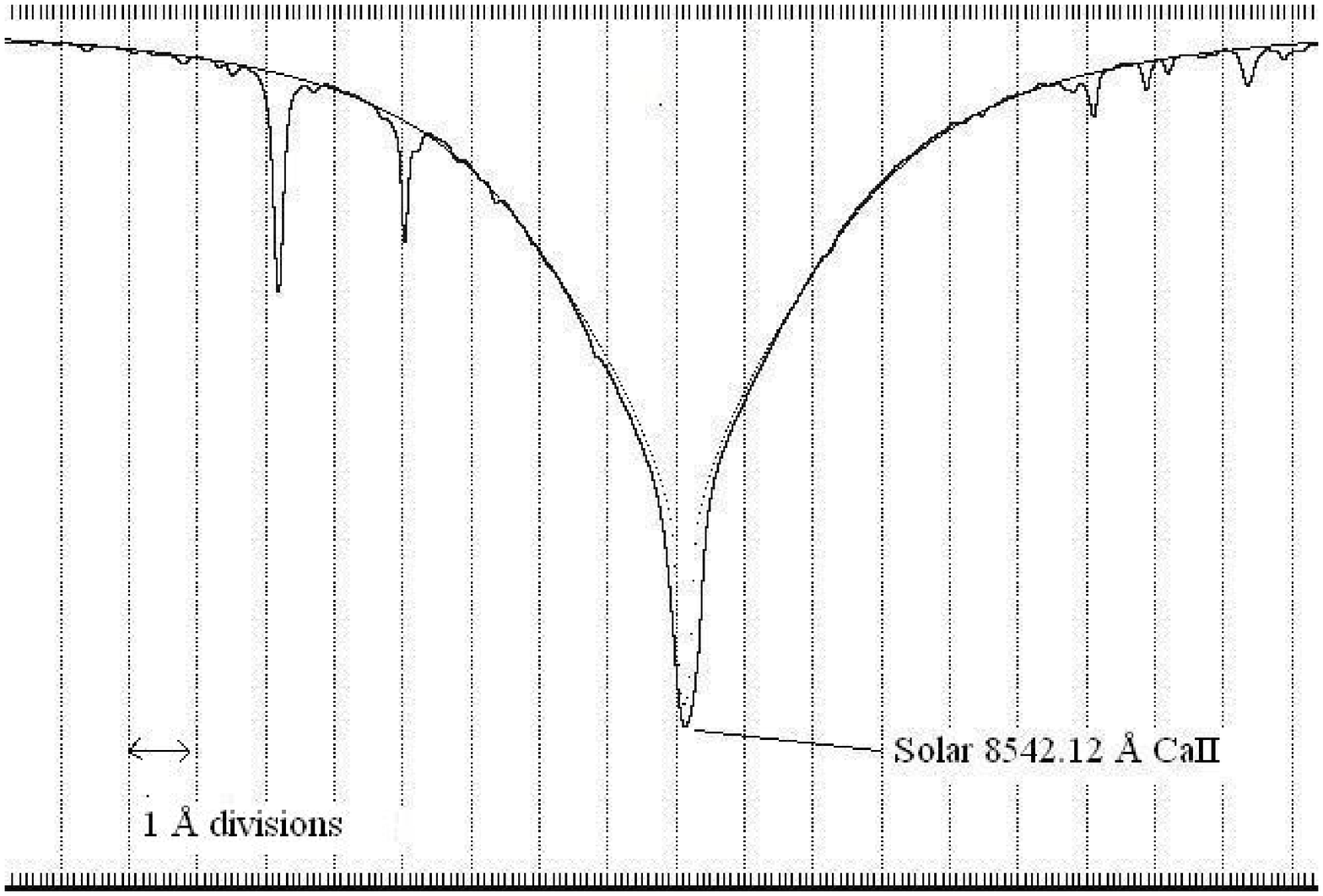}\end{flushleft}

\caption{{\footnotesize Fitting Model line profile to Observed Data for $\alpha$
Cen A and Sun to determine equivalent width and abundance for Strong
Ca}{\scriptsize II} {\footnotesize line 8542.120 $\AA$ \label{Fig-Fit for Aca&Sun}
(Scale: 1Å divisions)}}
\end{figure*}

As can be seen in Figure \ref{Fig-Fit for Aca&Sun} the right hand
wing of the line profile for the observed data for $\alpha$ Cen A
is higher than that of the synthesised line profile. This does not
happen in the Sun's line profile. The problem then is with the reduction
process. Care is taken to ensure that features like this are compensated
for in the determination of the abundances.

\subsection{Results}

The mean abundances are determined from 1 Mg, 7 Ca, 9 Si, 13 Ti, 6
Cr, and 22 Fe lines. Each element is represented by species that cover
weak neutral, weak ionised, and strong lines except for Mg that is
represented by one strong line. $\alpha$ Cen A shows a mean overabundance
of 0.12$\pm$0.06 dex. The error is calculated from the variations
in the individual species abundances.

These abundances are larger than those used for starting abundances
from Grevesse \& Sauval (1998). The individual equivalent widths ($W_{\lambda}$),
abundance (A), and $\Delta$ abundances ($\Delta$A) for $\alpha$
Cen A and the Sun are listed in Tables \ref{Tab-FinishedAbund1} and
\ref{Tab-FinishedAbund2}. The mean abundance ($\left\langle A\right\rangle $),
the $\Delta$ mean abundances ($\Delta$$\left\langle A\right\rangle $)
along with the starting abundances for $\alpha$ Cen A and the Sun
for each element and the final mean overabundance are listed in Table
\ref{Tab-Mean A & Del A}.

\begin{table*}

\caption{{\footnotesize Input Parameters and Abundances for $\alpha$ Cen
A and the Sun\label{Tab-FinishedAbund1}}}

\begin{flushleft}\begin{tabular}{ccccccccccc}
\hline 
{\footnotesize $\lambda$}&
&
&
&
\multicolumn{2}{c}{{\footnotesize VDW $^{(2)}$}}&
\multicolumn{2}{c}{{\footnotesize $\alpha$ Cen A $^{(1)}$}}&
\multicolumn{2}{c}{{\footnotesize Sun $^{(1)}$}}&
{\footnotesize $\Delta$$\left\langle A\right\rangle $$^{(1)}$ }\tabularnewline
\cline{1-1} \cline{2-2} \cline{3-3} \cline{4-4} \cline{11-11} 
\hline 
{\footnotesize ($\AA$)}&
{\footnotesize Species$^{(2)}$}&
{\footnotesize E$_{\textrm{low}}$$^{(2)}$}&
{\footnotesize log(gf)$^{(2)}$}&
{\footnotesize $\sigma$}&
{\footnotesize $\alpha$}&
{\footnotesize W$_{\lambda}$ (m$\AA$)}&
{\footnotesize A}&
{\footnotesize W$_{\lambda}$  (m$\AA$)}&
{\footnotesize A }&
{\footnotesize (dex)}\tabularnewline
\hline
{\footnotesize 4508.287}&
{\footnotesize Fe}{\scriptsize II}&
{\footnotesize 2.84}&
{\footnotesize -2.520}&
{\footnotesize 188}&
{\footnotesize 0.267}&
{\footnotesize 109.53}&
{\footnotesize 7.935}&
{\footnotesize 90.59}&
{\footnotesize 7.670}&
{\small 0.027}\tabularnewline
{\footnotesize 4602.006}&
{\footnotesize Fe}{\scriptsize I}&
{\footnotesize 1.61}&
{\footnotesize -3.150}&
{\footnotesize 296}&
{\footnotesize 0.260}&
{\footnotesize 91.08}&
{\footnotesize 7.870}&
{\footnotesize 73.68}&
{\footnotesize 7.680}&
{\footnotesize 0.190}\tabularnewline
{\footnotesize 4656.979}&
{\footnotesize Fe}{\scriptsize II}&
{\footnotesize 2.88}&
{\footnotesize -3.580}&
{\footnotesize 190}&
{\footnotesize 0.330}&
{\footnotesize 36.80}&
{\footnotesize 7.346}&
{\footnotesize 31.97}&
{\footnotesize 7.330}&
{\footnotesize 0.016}\tabularnewline
{\footnotesize 4758.122}&
{\footnotesize Ti}{\scriptsize I}&
{\footnotesize 2.25}&
{\footnotesize 0.481}&
{\footnotesize 326}&
{\footnotesize 0.246}&
{\footnotesize 54.95}&
{\footnotesize 5.027}&
{\footnotesize 43.46}&
{\footnotesize 4.950}&
{\footnotesize 0.077}\tabularnewline
{\footnotesize 4759.274}&
{\footnotesize Ti}{\scriptsize I}&
{\footnotesize 2.25}&
{\footnotesize 0.570}&
{\footnotesize 327}&
{\footnotesize 0.246}&
{\footnotesize 57.32}&
{\footnotesize 4.990}&
{\footnotesize 47.90}&
{\footnotesize 4.958}&
{\footnotesize 0.032}\tabularnewline
{\footnotesize 4798.535}&
{\footnotesize Ti}{\scriptsize II}&
{\footnotesize 1.08}&
{\footnotesize -2.670}&
{\footnotesize 211}&
{\footnotesize 0.209}&
{\footnotesize 54.95}&
{\footnotesize 5.130}&
{\footnotesize 41.77}&
{\footnotesize 5.000}&
{\footnotesize 0.130}\tabularnewline
{\footnotesize 4801.028}&
{\footnotesize Cr}{\scriptsize I}&
{\footnotesize 3.12}&
{\footnotesize -0.131}&
{\footnotesize 348}&
{\footnotesize 0.240}&
{\footnotesize 59.73}&
{\footnotesize 5.800}&
{\footnotesize 49.66}&
{\footnotesize 5.720}&
{\footnotesize 0.800}\tabularnewline
{\footnotesize 4964.931}&
{\footnotesize Cr}{\scriptsize I}&
{\footnotesize 0.94}&
{\footnotesize -2.527}&
{\footnotesize 262}&
{\footnotesize 0.291}&
{\footnotesize 49.04}&
{\footnotesize 5.790}&
{\footnotesize 35.90}&
{\footnotesize 5.720}&
{\footnotesize 0.700}\tabularnewline
{\footnotesize 5113.445}&
{\footnotesize Ti}{\scriptsize I}&
{\footnotesize 1.4}&
{\footnotesize -0.727}&
{\footnotesize 298}&
{\footnotesize 0.243}&
{\footnotesize 36.10}&
{\footnotesize 5.003}&
{\footnotesize 25.15}&
{\footnotesize 4.950}&
{\footnotesize 0.053}\tabularnewline
{\footnotesize 5225.533}&
{\footnotesize Fe}{\scriptsize I}&
{\footnotesize 0.11}&
{\footnotesize -4.790}&
{\footnotesize 207}&
{\footnotesize 0.253}&
{\footnotesize 86.83}&
{\footnotesize 7.823}&
{\footnotesize 71.07}&
{\footnotesize 7.680}&
{\footnotesize 0.143}\tabularnewline
{\footnotesize 5232.952}&
{\footnotesize Fe}{\scriptsize I}&
{\footnotesize 2.94}&
{\footnotesize -0.058}&
{\footnotesize 713}&
{\footnotesize 0.238}&
{\footnotesize 382.43}&
{\footnotesize 7.450}&
{\footnotesize 376.95}&
{\footnotesize 7.510}&
{\footnotesize -0.060}\tabularnewline
{\footnotesize 5234.632}&
{\footnotesize Fe}{\scriptsize II}&
{\footnotesize 3.21}&
{\footnotesize -2.230}&
{\footnotesize 188}&
{\footnotesize 0.268}&
{\footnotesize 105.42}&
{\footnotesize 7.813}&
{\footnotesize 89.23}&
{\footnotesize 7.570}&
{\footnotesize 0.243}\tabularnewline
{\footnotesize 5247.057}&
{\footnotesize Fe}{\scriptsize I}&
{\footnotesize 0.09}&
{\footnotesize -4.950}&
{\footnotesize 206}&
{\footnotesize 0.253}&
{\footnotesize 76.63}&
{\footnotesize 7.678}&
{\footnotesize 64.17}&
{\footnotesize 7.630}&
{\footnotesize 0.048}\tabularnewline
{\footnotesize 5272.002}&
{\footnotesize Cr}{\scriptsize I}&
{\footnotesize 3.45}&
{\footnotesize -0.422}&
{\footnotesize 757}&
{\footnotesize 0.238}&
{\footnotesize 35.43}&
{\footnotesize 5.820}&
{\footnotesize 21.92}&
{\footnotesize 5.630}&
{\footnotesize 0.190}\tabularnewline
{\footnotesize 5295.781}&
{\footnotesize Ti}{\scriptsize I}&
{\footnotesize 1.05}&
{\footnotesize -1.57}&
{\footnotesize 278}&
{\footnotesize 0.253}&
{\footnotesize 18.13}&
{\footnotesize 5.017}&
{\footnotesize 11.60}&
{\footnotesize 4.990}&
{\footnotesize 0.027}\tabularnewline
{\footnotesize 5300.751}&
{\footnotesize Cr}{\scriptsize I}&
{\footnotesize 0.98}&
{\footnotesize -2.129}&
{\footnotesize 329}&
{\footnotesize 0.263}&
{\footnotesize 71.50}&
{\footnotesize 5.890}&
{\footnotesize 54.71}&
{\footnotesize 5.720}&
{\footnotesize 0.170}\tabularnewline
{\footnotesize 5312.859}&
{\footnotesize Cr}{\scriptsize I}&
{\footnotesize 3.45}&
{\footnotesize -0.562}&
{\footnotesize 751}&
{\footnotesize 0.238}&
{\footnotesize 27.21}&
{\footnotesize 5.776}&
{\footnotesize 18.75}&
{\footnotesize 5.680}&
{\footnotesize 0.96}\tabularnewline
{\footnotesize 5336.793}&
{\footnotesize Ti}{\scriptsize II}&
{\footnotesize 1.58}&
{\footnotesize -1.630}&
{\footnotesize 272}&
{\footnotesize 0.314}&
{\footnotesize 90.84}&
{\footnotesize 5.300}&
{\footnotesize 73.21}&
{\footnotesize 5.050}&
{\footnotesize 0.250}\tabularnewline
{\footnotesize 5418.773}&
{\footnotesize Ti}{\scriptsize II}&
{\footnotesize 1.58}&
{\footnotesize -2.110}&
{\footnotesize 270}&
{\footnotesize 0.315}&
{\footnotesize 61.42}&
{\footnotesize 5.140}&
{\footnotesize 47.55}&
{\footnotesize 4.980}&
{\footnotesize 0.160}\tabularnewline
{\footnotesize 5490.154}&
{\footnotesize Ti}{\scriptsize I}&
{\footnotesize 1.46}&
{\footnotesize -0.877}&
{\footnotesize 374}&
{\footnotesize 0.262}&
{\footnotesize 28.78}&
{\footnotesize 4.984}&
{\footnotesize 19.64}&
{\footnotesize 4.945}&
{\footnotesize 0.39}\tabularnewline
{\footnotesize 5665.557}&
{\footnotesize Si}{\scriptsize I}&
{\footnotesize 4.92}&
{\footnotesize -1.940}&
{\footnotesize 1772}&
{\footnotesize 0.222}&
{\footnotesize 56.34}&
{\footnotesize 7.680}&
{\footnotesize 39.12}&
{\footnotesize 7.480}&
{\footnotesize 0.200}\tabularnewline
{\footnotesize 5684.490}&
{\footnotesize Si}{\scriptsize I}&
{\footnotesize 4.95}&
{\footnotesize -1.550}&
{\footnotesize 1798}&
{\footnotesize 0.221}&
{\footnotesize 77.74}&
{\footnotesize 7.610}&
{\footnotesize 66.87}&
{\footnotesize 7.500}&
{\footnotesize 0.110}\tabularnewline
{\footnotesize 5690.425}&
{\footnotesize Si}{\scriptsize I}&
{\footnotesize 4.93}&
{\footnotesize -1.770}&
{\footnotesize 1772}&
{\footnotesize 0.222}&
{\footnotesize 62.79}&
{\footnotesize 7.616}&
{\footnotesize 52.84}&
{\footnotesize 7.520}&
{\footnotesize 0.096}\tabularnewline
{\footnotesize 5701.106}&
{\footnotesize Si}{\scriptsize I}&
{\footnotesize 4.93}&
{\footnotesize -1.950}&
{\footnotesize 1768}&
{\footnotesize 0.222}&
{\footnotesize 55.19}&
{\footnotesize 7.680}&
{\footnotesize 39.91}&
{\footnotesize 7.510}&
{\footnotesize 0.170}\tabularnewline
{\footnotesize 5708.402}&
{\footnotesize Si}{\scriptsize I}&
{\footnotesize 4.95}&
{\footnotesize -1.370}&
{\footnotesize 1787}&
{\footnotesize 0.222}&
{\footnotesize 95.29}&
{\footnotesize 7.640}&
{\footnotesize 84.96}&
{\footnotesize 7.537}&
{\footnotesize 0.103}\tabularnewline
{\footnotesize 5787.922}&
{\footnotesize Cr}{\scriptsize I}&
{\footnotesize 3.32}&
{\footnotesize -0.083}&
{\footnotesize 1097}&
{\footnotesize 0.291}&
{\footnotesize 87.12}&
{\footnotesize 6.203}&
{\footnotesize 48.17}&
{\footnotesize 5.660}&
{\footnotesize 0.543}\tabularnewline
{\footnotesize 5866.457}&
{\footnotesize Ti}{\scriptsize I}&
{\footnotesize 1.07}&
{\footnotesize -0.784}&
{\footnotesize 259}&
{\footnotesize 0.262}&
{\footnotesize 59.83}&
{\footnotesize 5.120}&
{\footnotesize 43.38}&
{\footnotesize 4.980}&
{\footnotesize 0.140}\tabularnewline
{\footnotesize 5916.254}&
{\footnotesize Fe}{\scriptsize I}&
{\footnotesize 2.45}&
{\footnotesize -2.990}&
{\footnotesize 341}&
{\footnotesize 0.238}&
{\footnotesize 64.02}&
{\footnotesize 7.710}&
{\footnotesize 52.80}&
{\footnotesize 7.650}&
{\footnotesize 0.060}\tabularnewline
{\footnotesize 5922.115}&
{\footnotesize Ti}{\scriptsize I}&
{\footnotesize 1.05}&
{\footnotesize -1.410}&
{\footnotesize 313}&
{\footnotesize 0.242}&
{\footnotesize 28.10}&
{\footnotesize 5.065}&
{\footnotesize 17.82}&
{\footnotesize 5.000}&
{\footnotesize 0.065}\tabularnewline
{\footnotesize 5948.545}&
{\footnotesize Si}{\scriptsize I}&
{\footnotesize 5.08}&
{\footnotesize -1.130}&
{\footnotesize 1875}&
{\footnotesize 0.222}&
{\footnotesize 107.51}&
{\footnotesize 7.663}&
{\footnotesize 94.74}&
{\footnotesize 7.506}&
{\footnotesize 0.157}\tabularnewline
{\footnotesize 6082.715}&
{\footnotesize Fe}{\scriptsize I}&
{\footnotesize 2.22}&
{\footnotesize -3.570}&
{\footnotesize 306}&
{\footnotesize 0.271}&
{\footnotesize 47.38}&
{\footnotesize 7.697}&
{\footnotesize 33.24}&
{\footnotesize 7.590}&
{\footnotesize 0.107}\tabularnewline
{\footnotesize 6092.799}&
{\footnotesize Ti}{\scriptsize I}&
{\footnotesize 1.89}&
{\footnotesize -1.323}&
{\footnotesize 398}&
{\footnotesize 0.239}&
{\footnotesize 7.92}&
{\footnotesize 5.125}&
{\footnotesize 3.77}&
{\footnotesize 4.96}&
{\footnotesize 0.165}\tabularnewline
{\footnotesize 6151.623}&
{\footnotesize Fe}{\scriptsize I}&
{\footnotesize 2.18}&
{\footnotesize -3.300}&
{\footnotesize 277}&
{\footnotesize 0.263}&
{\footnotesize 62.73}&
{\footnotesize 7.711}&
{\footnotesize 49.26}&
{\footnotesize 7.61}&
{\footnotesize 0.101}\tabularnewline
{\footnotesize 6161.297}&
{\footnotesize Ca}{\scriptsize I}&
{\footnotesize 2.52}&
{\footnotesize -1.266}&
{\footnotesize 978}&
{\footnotesize 0.257}&
{\footnotesize 79.49}&
{\footnotesize 6.540}&
{\footnotesize 66.02}&
{\footnotesize 6.409}&
{\footnotesize 0.131}\tabularnewline
{\footnotesize 6162.183}&
{\footnotesize Ca}{\scriptsize I}&
{\footnotesize 1.89}&
{\footnotesize -0.097}&
{\footnotesize 878}&
{\footnotesize 0.236}&
{\footnotesize 280.96}&
{\footnotesize 6.350}&
{\footnotesize 277.82}&
{\footnotesize 6.350}&
{\footnotesize 0}\tabularnewline
{\footnotesize 6166.441}&
{\footnotesize Ca}{\scriptsize I}&
{\footnotesize 2.52}&
{\footnotesize -1.142}&
{\footnotesize 976}&
{\footnotesize 0.257}&
{\footnotesize 81.87}&
{\footnotesize 6.454}&
{\footnotesize 75.14}&
{\footnotesize 6.420}&
{\footnotesize 0.034}\tabularnewline
{\footnotesize 6173.342}&
{\footnotesize Fe}{\scriptsize I}&
{\footnotesize 2.2}&
{\footnotesize -2.880}&
{\footnotesize 281}&
{\footnotesize 0.266}&
{\footnotesize 81.34}&
{\footnotesize 7.740}&
{\footnotesize 70.01}&
{\footnotesize 7.664}&
{\footnotesize 0.076}\tabularnewline
{\footnotesize 6200.321}&
{\footnotesize Fe}{\scriptsize I}&
{\footnotesize 2.61}&
{\footnotesize -2.440}&
{\footnotesize 350}&
{\footnotesize 0.235}&
{\footnotesize 85.81}&
{\footnotesize 7.750}&
{\footnotesize 74.75}&
{\footnotesize 7.680}&
{\footnotesize 0.070}\tabularnewline
{\footnotesize 6258.109}&
{\footnotesize Ti}{\scriptsize I}&
{\footnotesize 1.44}&
{\footnotesize -0.299}&
{\footnotesize 355}&
{\footnotesize 0.237}&
{\footnotesize 62.19}&
{\footnotesize 5.010}&
{\footnotesize 51.87}&
{\footnotesize 4.980}&
{\footnotesize 0.030}\tabularnewline
{\footnotesize 6297.801}&
{\footnotesize Fe}{\scriptsize I}&
{\footnotesize 2.2}&
{\footnotesize -2.750}&
{\footnotesize 278}&
{\footnotesize 0.264}&
{\footnotesize 86.68}&
{\footnotesize 7.700}&
{\footnotesize 76.60}&
{\footnotesize 7.650}&
{\footnotesize 0.050}\tabularnewline
\hline
\end{tabular}\end{flushleft}

{\footnotesize $^{(\textrm{1})}$This paper; $^{(\textrm{2})}$Allende
Prieto et al. 2001}
\end{table*}

\begin{table*}

\caption{{\footnotesize Input Parameters and Abundances for $\alpha$ Cen
A and the Sun (Continued)\label{Tab-FinishedAbund2}}}

\begin{flushleft}\begin{tabular}{ccccccccccc}
\hline 
{\footnotesize $\lambda$}&
&
&
&
\multicolumn{2}{c}{{\footnotesize VDW$^{(2)}$}}&
\multicolumn{2}{c}{{\footnotesize $\alpha$ Cen A$^{(1)}$}}&
\multicolumn{2}{c}{{\footnotesize Sun$^{(1)}$}}&
{\footnotesize $\Delta$$\left\langle A\right\rangle $$^{(1)}$ }\tabularnewline
\cline{1-1} \cline{2-2} \cline{3-3} \cline{4-4} \cline{11-11} 
\hline 
{\footnotesize ($\AA$)}&
{\footnotesize Species$^{(2)}$}&
{\footnotesize E$_{\textrm{low}}$$^{(2)}$}&
{\footnotesize log(gf)$^{(2)}$}&
{\footnotesize $\sigma$}&
{\footnotesize $\alpha$}&
{\footnotesize W$_{\lambda}$  (m$\AA$)}&
{\footnotesize A}&
{\footnotesize W$_{\lambda}$  (m$\AA$)}&
{\footnotesize A}&
{\footnotesize (dex)}\tabularnewline
\hline
{\footnotesize 6371.361}&
{\footnotesize Si}{\scriptsize II}&
{\footnotesize 8.12}&
{\footnotesize -0.000}&
{\footnotesize 389}&
{\footnotesize 0.189}&
{\footnotesize 43.34}&
{\footnotesize 7.850}&
{\footnotesize 35.22}&
{\footnotesize 7.520}&
{\footnotesize 0.330}\tabularnewline
{\footnotesize 6432.684}&
{\footnotesize Fe}{\scriptsize II}&
{\footnotesize 2.89}&
{\footnotesize -2.510}&
{\footnotesize 174}&
{\footnotesize 0.270}&
{\footnotesize 53.76}&
{\footnotesize 7.600}&
{\footnotesize 43.42}&
{\footnotesize 7.449}&
{\footnotesize 0.151}\tabularnewline
{\footnotesize 6455.604}&
{\footnotesize Ca}{\scriptsize I}&
{\footnotesize 2.52}&
{\footnotesize -1.290}&
{\footnotesize 365}&
{\footnotesize 0.241}&
{\footnotesize 69.90}&
{\footnotesize 6.490}&
{\footnotesize 56.85}&
{\footnotesize 6.350}&
{\footnotesize 0.140}\tabularnewline
{\footnotesize 6481.878}&
{\footnotesize Fe}{\scriptsize I}&
{\footnotesize 2.28}&
{\footnotesize -2.980}&
{\footnotesize 308}&
{\footnotesize 0.243}&
{\footnotesize 76.78}&
{\footnotesize 7.757}&
{\footnotesize 65.20}&
{\footnotesize 7.680}&
{\footnotesize 0.077}\tabularnewline
{\footnotesize 6498.945}&
{\footnotesize Fe}{\scriptsize I}&
{\footnotesize 0.96}&
{\footnotesize -4.700}&
{\footnotesize 226}&
{\footnotesize 0.253}&
{\footnotesize 57.82}&
{\footnotesize 7.740}&
{\footnotesize 43.20}&
{\footnotesize 7.658}&
{\footnotesize 0.082}\tabularnewline
{\footnotesize 6499.656}&
{\footnotesize Ca}{\scriptsize I}&
{\footnotesize 2.52}&
{\footnotesize -0.818}&
{\footnotesize 364}&
{\footnotesize 0.239}&
{\footnotesize 101.06}&
{\footnotesize 6.600}&
{\footnotesize 92.17}&
{\footnotesize 6.500}&
{\footnotesize 0.100}\tabularnewline
{\footnotesize 6516.086}&
{\footnotesize Fe}{\scriptsize II}&
{\footnotesize 2.89}&
{\footnotesize -3.380}&
{\footnotesize 174}&
{\footnotesize 0.270}&
{\footnotesize 45.07}&
{\footnotesize 7.270}&
{\footnotesize 53.99}&
{\footnotesize 7.550}&
{\footnotesize -0.280}\tabularnewline
{\footnotesize 6750.161}&
{\footnotesize Fe}{\scriptsize I}&
{\footnotesize 2.42}&
{\footnotesize -2.620}&
{\footnotesize 335}&
{\footnotesize 0.241}&
{\footnotesize 111.89}&
{\footnotesize 8.150}&
{\footnotesize 79.07}&
{\footnotesize 7.695}&
{\footnotesize 0.455}\tabularnewline
{\footnotesize 6978.861}&
{\footnotesize Fe}{\scriptsize I}&
{\footnotesize 2.48}&
{\footnotesize -2.500}&
{\footnotesize 337}&
{\footnotesize 0.241}&
{\footnotesize 108.08}&
{\footnotesize 8.000}&
{\footnotesize 82.04}&
{\footnotesize 7.662}&
{\footnotesize 0.338}\tabularnewline
{\footnotesize 7357.735}&
{\footnotesize Ti}{\scriptsize I}&
{\footnotesize 1.44}&
{\footnotesize -1.06}&
{\footnotesize 329}&
{\footnotesize 0.244}&
{\footnotesize 29.57}&
{\footnotesize 5.060}&
{\footnotesize 20.77}&
{\footnotesize 5.030}&
{\footnotesize 0.030}\tabularnewline
{\footnotesize 7515.836}&
{\footnotesize Fe}{\scriptsize II}&
{\footnotesize 3.9}&
{\footnotesize -3.450}&
{\footnotesize 187}&
{\footnotesize 0.271}&
{\footnotesize 23.33}&
{\footnotesize 7.750}&
{\footnotesize 14.73}&
{\footnotesize 7.517}&
{\footnotesize 0.233}\tabularnewline
{\footnotesize 7680.271}&
{\footnotesize Si}{\scriptsize I}&
{\footnotesize 5.86}&
{\footnotesize -0.590}&
{\footnotesize 2107}&
{\footnotesize 0.495}&
{\footnotesize 101.18}&
{\footnotesize 7.640}&
{\footnotesize 101.78}&
{\footnotesize 7.610}&
{\footnotesize 0.030}\tabularnewline
{\footnotesize 7711.730}&
{\footnotesize Fe}{\scriptsize II}&
{\footnotesize 3.9}&
{\footnotesize -2.450}&
{\footnotesize 186}&
{\footnotesize 0.264}&
{\footnotesize 58.42}&
{\footnotesize 7.560}&
{\footnotesize 50.79}&
{\footnotesize 7.430}&
{\footnotesize 0.130}\tabularnewline
{\footnotesize 7918.387}&
{\footnotesize Si}{\scriptsize I}&
{\footnotesize 5.95}&
{\footnotesize -0.510}&
{\footnotesize 2934}&
{\footnotesize 0.232}&
{\footnotesize 108.85}&
{\footnotesize 7.600}&
{\footnotesize 102.00}&
{\footnotesize 7.504}&
{\footnotesize 0.096}\tabularnewline
{\footnotesize 8327.067}&
{\footnotesize Fe}{\scriptsize I}&
{\footnotesize 2.20}&
{\footnotesize -1.525}&
{\footnotesize 258}&
{\footnotesize 0.247}&
{\footnotesize 226.21}&
{\footnotesize 7.660}&
{\footnotesize 198.38}&
{\footnotesize 7.621}&
{\footnotesize 0.039}\tabularnewline
{\footnotesize 8542.120}&
{\footnotesize Ca}{\scriptsize II}&
{\footnotesize 1.7}&
{\footnotesize -0.463}&
{\footnotesize 291}&
{\footnotesize 0.275}&
{\footnotesize 3206.68}&
{\footnotesize 6.360}&
{\footnotesize 3274.47}&
{\footnotesize 6.360}&
{\footnotesize 0}\tabularnewline
{\footnotesize 8662.169}&
{\footnotesize Ca}{\scriptsize II}&
{\footnotesize 1.69}&
{\footnotesize -0.723}&
{\footnotesize 291}&
{\footnotesize 0.275}&
{\footnotesize 2568.95}&
{\footnotesize 6.420}&
{\footnotesize 2566.40}&
{\footnotesize 6.380}&
{\footnotesize 0.040}\tabularnewline
{\footnotesize 8806.778}&
{\footnotesize Mg}{\scriptsize I}&
{\footnotesize 4.33}&
{\footnotesize -0.120}&
{\footnotesize 531}&
{\footnotesize 0.292}&
{\footnotesize 679.65}&
{\footnotesize 7.900}&
{\footnotesize 562.10}&
{\footnotesize 7.680}&
{\footnotesize 0.220}\tabularnewline
\hline
\end{tabular}\end{flushleft}

{\footnotesize $^{(\textrm{1})}$This paper; $^{(\textrm{2})}$Allende
Prieto et al. 2001}
\end{table*}

\begin{table*}

\caption{{\footnotesize Mean Abundances and Mean Overabundances for $\alpha$
Cen A models (AK), and Sun Model (SH)\label{Tab-Mean A & Del A}}}

\begin{flushleft}{\footnotesize }\begin{tabular}{ccccc}
\hline 
{\footnotesize Element}&
{\footnotesize Grevesse \& Sauval 1998}&
{\footnotesize $\alpha$ Cen A - $\left\langle A\right\rangle $ }&
{\footnotesize Solar - $\left\langle A\right\rangle $ }&
{\footnotesize $\Delta$$\left\langle A\right\rangle $ (dex)}\tabularnewline
\hline
{\footnotesize Mg}&
{\footnotesize 7.58$\pm$0.05}&
{\footnotesize 7.9 (1 line)}&
{\footnotesize 7.6800}&
{\footnotesize 0.200}\tabularnewline
{\footnotesize Ca}&
{\footnotesize 6.36$\pm$0.02}&
{\footnotesize 6.459$\pm$0.092}&
{\footnotesize 6.396$\pm0.054$}&
{\footnotesize 0.064$\pm$0.017}\tabularnewline
{\footnotesize Si}&
{\footnotesize 7.55$\pm$0.05}&
{\footnotesize 7.664$\pm$0.0756}&
{\footnotesize 7.521$\pm0.0370$}&
{\footnotesize 0.144$\pm$0.011}\tabularnewline
{\footnotesize Ti}&
{\footnotesize 5.02$\pm$0.06}&
{\footnotesize 5.075$\pm$0.088}&
{\footnotesize 4.983$\pm0.032$}&
{\footnotesize 0.092$\pm$0.019}\tabularnewline
{\footnotesize Cr}&
{\footnotesize 5.67$\pm$0.03}&
{\footnotesize 5.880$\pm$0.163}&
{\footnotesize 5.688$\pm$0.038}&
{\footnotesize 0.192$\pm$0.029}\tabularnewline
{\footnotesize Fe}&
{\footnotesize 7.50$\pm$0.05}&
{\footnotesize 7.714$\pm$0.197}&
{\footnotesize 7.599$\pm0.098$}&
{\footnotesize 0.115$\pm$0.029}\tabularnewline
\hline 
\multicolumn{5}{c}{{\footnotesize $\alpha$ Cen A's mean overabundance = 0.12$\pm$0.06
dex}}\tabularnewline
\hline
\end{tabular}\end{flushleft}
\end{table*}

\section{Summary and Discussions}

The mean abundance for the six elements investigated in the chemical
composition of $\alpha$ Cen A are Mg=7.9 %
\footnote{Only one line is analysed hence no error value.%
}, Ca=6.46$\pm$0.09, Si=7.66$\pm$0.07, Ti=5.07$\pm$0.09, Cr=5.88$\pm$0.16,
and Fe=7.71$\pm$0.20 dex. This leads to a mean abundance of 0.12$\pm$0.06
dex with respect to the Sun. 

Previous studies are not able to use strong lines as no reliable theory
existed to calculate the collisional broadening of these lines. For
this project the development of the ABO theory (Barklem et al. 1998a)
to calculate the VDW damping, enable us to use strong lines in determining
the mean abundance of $\alpha$ Cen A compared to the Sun.

The $\alpha$ Cen A parameters used are $T_{eff}$=5784.3$\pm$5.5K
and log($g_{s}$)=4.28$\pm$0.01. Two models for $\alpha$ Cen A are
used, the second one (AH) a scaled solar model for comparison with
the first (AK) to verify the validity of using an interpolated Kurucz
model (Kurucz 1979). Once the use of model AK is validated, this model
is used to synthesize line profiles for $\alpha$ Cen A to match the
observed line profiles.

Solar abundances are determined from comparing observed line profiles
from the Jungfraujoch Atlas with those of the Holweger-M\"uller solar
model (Holweger \& M\"uller 1974)).

The results of this study, that $\alpha$ Cen A is overabundant with
respect to the Sun and can be included with other metal rich stars,
agrees with those of previous studies (Noels et al. 1991; Chmielewski
et al. 1992; Neuforge 1993; Neuforge-Verheecke \& Magain 1997) with
an exact agreement of mean overabundance of 0.12$\pm$0.06 dex by
Furenlid \& Meylan (1990). This mean overabundance indicates that
$\alpha$ Cen A did not originate in the same cloud as the Sun but
from material that is more enriched by stellar nuclear processing.

The above studies did not use strong lines or the ABO theory. Our
result supports the determination that reliable abundances can be
derived from strong lines provided the ABO theory is used to calculate
the VDW damping.

By using the ABO theory for strong lines, the analysis of spectra,
construction of model atmospheres, and the subsequent synthesised
line profiles, the chemical composition of apparently faint stars,
such as those in external galaxies can be determined. All galaxies
contain cool F,G K type stars whose spectra contain strong metallic
lines. Previously these lines were not able to be used for absolute
abundances due uncertainties in the theory for calculating the VDW
damping. With the ABO theory now firmly established and reliable model
atmospheres existing for cool stars it will be possible to extend
reliable abundance analyses to more and more distant galaxies using
strong lines.

\subsection*{Acknowledgments}

I would like to acknowledge and thank Paul Barklem for the use of
the VDW damping values calculated from the ABO theory and the Astrophysics
Group at the University of Queensland. I would also like to thank
the anonymous referees for their constructive comments.


\section*{References}

\reference Ahrens T.J., ed.\  1995, Global Earth Physics, a Handbook of Physical
Constants

\reference Allende Prieto C., 
Barklem P.S., Asplund M., Ruiz Cobo B., 2001, ApJ, 558, 830

\reference Anstee S.D., 
O'Mara B.J., 1991, MNRAS, 253, 549

\reference Anstee S.D., 
O'Mara B.J, 1995, MNRAS, 276, 859

\reference Barklem P.S., 
Anstee S,D., O'Mara B.J., 1998a, PASA, 15, 336

\reference Barklem P.S., 
O'Mara B.J., 1997, MNRAS, 290, 102

\reference Barklem P.S., 
O'Mara B.J., Ross J.E., 1998b, MNRAS, 296, 1057

\reference Bessell M. S., 
Castelli F., Plez B., 1998, AA, 333, 231

\reference Blackwell D.E, 
Shallis M.J., 1977, MNRAS, 180, 177

\reference Brueckner K.A., 1971, ApJ, 169, 621

\reference Chmielewski Y., 
Friel E., Cayrel de Strobel G., Bentolila C., 1992, AA, 263, 219

\reference Delbouille L., 
Roland C., 1995, ASP, 81,32

\reference Demarque P., 
Guenther D.B., van Altena W.F., 1986, ApJ, 300, 773

\reference Furenlid I., 
Meylan T., 1990, ApJ, 350, 827

\reference Grevesse N., 
Sauval A.J., 1998, Space Science Review, 85, 161

\reference Holweger H., 
M\"uller E.A., 1974, Solar Physics. 39, 19

\reference Kurucz R.L., 1979, ApJS, 40, 1

\reference Neuforge C., 1993, AA, 268, 650

\reference Neuforge-Verheecke C., 
Magain P., 1997, AA, 328, 261

\reference Noels A., 
Grevesse N., Magain P., Neuforge C., Baglin A., Lebreton Y., 1991, AA, 247, 91

\reference O'Mara B.J., 1976, MNRAS, 177, 551

\reference Perryman M.A.C., 
ESA, 1997, The HIPPARCOS and TYCHO Catalogues. Astrometric
and Photometric catalogues derived from the ESA Hipparcos Space Astrometry
Mission, ESA Publications Division, Netherlands, 1997, vol 1200

\reference Soderblom D.R., 1986, AA, 158, 273






\end{document}